\documentclass[fleqn,usenatbib]{mnras}
\usepackage{mathptmx}
\usepackage{txfonts}
\usepackage{hyperref}
\usepackage{academicons}
\usepackage{xcolor}

\newcommand{\orcid}[1]{\href{https://orcid.org/#1}{\textcolor[HTML]{A6CE39}{\aiOrcid}}}

\usepackage[T1]{fontenc}

\DeclareRobustCommand{\VAN}[3]{#2}
\let\VANthebibliography\thebibliography
\def\thebibliography{\DeclareRobustCommand{\VAN}[3]{##3}\VANthebibliography}

\usepackage{graphicx}	

\title[Herschel\,36 orbit with VLTI]{The outer orbit of the high-mass stellar triple system Herschel\,36 determined with the VLTI}

\author[J. Sanchez-Bermudez, et al.]{
J. Sanchez-Bermudez,$^{1,2}$\thanks{E-mail: joelsb@astro.unam.mx}
 C.~A. Hummel,$^{3}$
 J. D\'iaz-L\'opez,$^{1}$
A. Alberdi,$^{4}$
R. Sch\"odel,$^{4}$
 J. I. Arias,$^{5}$
 \newauthor
 R. H. Barb\'a\thanks{Deceased},$^{5}$
E. Bastida-Escamilla,$^{6}$
W. Brandner,$^{2}$
J. Ma\'iz Apell\'aniz,$^{3,7}$
and  J.-U. Pott,$^{2}$
\\
$^{1}$Instituto de Astronom\'ia, Universidad Nacional Aut\'onoma de M\'exico, Apdo. Postal 70264, Ciudad de M\'exico, 04510, M\'exico\\
$^{2}$Max-Planck-Institut f\"ur Astronomie, K\"{o}nigstuhl 17, D-69117 Heidelberg, Germany\\
$^{3}$European Southern Observatory, Karl-Schwarzschild-strasse 2, D-85748 Garching, Germany\\
$^{4}$Instituto de Astrof\'isica de Andaluc\'ia (IAA-CSIC), Glorieta de la Astronom\'ia S/N, 18008, Granada, Spain\\
$^{5}$Departamento de F\'isica y Astronom\'ia, Universidad de La Serena, Av. Cisternas 1200 Norte, La Serena, Chile\\
$^{6}$Tecnol\'ogico de Monterrey, Escuela de Ingenier\'ia y Ciencias, Ave. Eugenio Garza Sada 2501, Monterrey, N.L., Mexico, 64849\\
$^{7}$Centro de Astrobiolog\'ia, CSIC-INTA, Campus ESAC Camino bajo del castillo s/n, E-28692 Villanueva de la Ca\~nada, Spain\
}

\date{Accepted XXX. Received YYY; in original form ZZZ}

\pubyear{2022}

\begin{document}
\label{firstpage}
\pagerange{\pageref{firstpage}--\pageref{lastpage}}
\maketitle

\begin{abstract}
Multiplicity is a ubiquitous characteristic of massive stars. Multiple systems offer us a unique observational constraint on the formation of high-mass systems. Herschel\,36\,A is a massive triple system composed of a close binary (Ab1-Ab2) and an outer component (Aa). We measured the orbital motion of the outer component of Herschel\,36\,A using infrared interferometry with the AMBER and PIONIER instruments of ESO's Very Large Telescope Interferometer. Our immediate aims are to constrain the masses of all components of this system and to determine if the outer orbit is co-planar with the inner one. Reported spectroscopic data for all three components of this system and our interferometric data allow us to derive full orbital solutions for the outer orbit Aa-Ab and the inner orbit Ab1-Ab2. For the first time, we derive the absolute masses of m$_{\mathrm{Aa}}$=22.3 $\pm$ 1.7M$_{\odot}$, m$_{\mathrm{Ab1}}$ = 20.5 $\pm$ 1.5 M$_{\odot}$ and m$_{\mathrm{Ab2}}$ = 12.5 $\pm$ 0.9 M$_{\odot}$. Despite not being able to resolve the close binary components, we infer the inclination of their orbit by imposing the same parallax as the outer orbit. Inclinations derived from the inner and outer orbits imply a modest difference of about 22$^{\circ}$ between the orbital planes. We discuss this result and the formation of Herschel 36\,A in the context of \textit{Core Accretion} and \textit{Competitive Accretion} models, which make different predictions regarding the statistic of the relative orbital inclinations.
\end{abstract}




\section{Introduction}

One of the most important characteristics to explain in the formation of massive stars is multiplicity. It is now well-known that at least 90\% of the O-stars possess at least one massive stellar companion \citep{Sana_2014, Sota_2014}. There are several physical properties that strongly depend on multiplicity \citep{Sana_2011}, for example: (a) the different evolutionary paths of massive multiples versus single stars \citep{Langer_2008, Crowther_2010} ; (b) the role of winds in the creation of dust in evolved systems \citep{Tuthill_2008}; or (c) the velocity dispersion of massive clusters \citep{Gieles_2010}.   
 
Properly characterizing dynamical interactions between different stellar components is important to determine the initial conditions of the formation process and to discern between different physical scenarios, for example: disk or filament fragmentation \citep{Bonnell_1992, Monin_2007}, stellar collisions and mergers \citep{Zinnecker_2002, Bonnell_2005}, or disk-assisted capture \citep{Bally_2005}. In order to do this, multi-epoch studies combining different techniques (such as interferometry, spectroscopy or adaptive-optics imaging) are necessary to study the geometries of massive multiples and their correlation with the proposed models. 

Different predictions can be proposed depending on the used model. On the one hand, the disk fragmentation model suggest that stellar companions are formed from instabilities in the accretion disks \citep[see e.g., the simulations in ][]{Krumholz_2007}. Thus, this scenario suggest that several formed stellar companions would remain orbiting the central source following coplanar orbits, preserving the original angular momentum of the fragmenting disk. On the other hand, in competitive accretion, the formation of massive stars depends on the reservoir of material from the "large-scale" environment. This scenario not only explains the formation of massive stars but of a entire initial-mass-function in a forming cluster. The most massive stars are formed at loci of the cloud with the strongest gravitational potential, which lead them to gain more material than the less massive stars. This scenario describes a dynamical environment, with mass segregation during the formation process \citep{Bonnell_2006, Bonnell_2007}. Massive multiples, thus, could be formed via dynamical interactions. This condition does not favour coplanar orbits of forming massive multiples, but more randomly oriented orbits.

The target of this study is Herschel\,36\,A, an intriguing hierarchical triple system with a combined luminosity that matches the theoretical luminosity of three ZAMS stars \citep{Arias_2010}, suggesting that the system is in a very early evolutionary stage with an age of the order of $\sim$1 Ma. The source is located at 1234$\pm$16 pc  \citep{Maiz-Apellaniz_2022} and it is responsible for the ionization of the central part of the M8 nebula. Herschel\,36\,A consists of three known components. Two of them, Ab1 (O9.5 V) - Ab2 (B0.7 V), form a close binary with a period of the circular orbit of 1.54157 days, while the third one, Aa (O7.5 Vz), moves on a wider eccentric orbit \citep[with a period of 492.8 days and an eccentricity of 0.29;][]{Campillay_2019}. 
 
In 2014, \citet{Sanchez-Bermudez_2014b} observed the source with AMBER  \citep{AMBER_Petrov_2007} at the Very Large Telescope Interferometer (VLTI), resolving, for the first time, the tertiary component. These observations also show that Aa is as bright as the combined Ab1-Ab2 pair. This result is interesting, since in hierarchical triple systems the most massive and luminous object usually forms part of the inner binary \citep[see e.g., ][]{Sanchez-Bermudez_2013, Mahy_2018}. Therefore, this system deserves a particular study to compare its current configuration along with its extreme youth with plausible models of star formation. 
 
This work presents the results of our monitoring program with the VLTI instruments to trace the orbit of Aa around the Ab1-Ab2 system. The paper is organized as follows: Sect.\,\ref{Sec:observations} presents our observations and data reduction. In Sect.\,\ref{Sec:analysis}, the analysis of the interferometric observables and of the orbit of the system Aa+Ab is presented, followed by a discussion in Sect.\,\ref{Sec:discussion}. In Sect.\,\ref{sec:summary}, we present a summary of this work. 

\begin{table*}
\caption[]{Herschel\,36\,A interferometric PIONIER ($H-$band) and AMBER ($K-$band) observations}
\label{tab:Obs}
\centering
\begin{tabular}{l c c c c c}     
\hline
                & Apr. 2014          & Sep. 2014          &  Aug. 2017         & Apr.2018           & Aug. 2018  \\
\hline
Array           & UT1+UT2+UT4        & D0+G1+H0+I1        & UT1+UT3+UT4        & UT1+UT3+UT4        & UT1+UT3+UT4 \\
Beam size [mas] & 7.66 $\times$ 1.57 & 3.00 $\times$ 2.56 & 4.70 $\times$ 1.81 & 6.40 $\times$ 1.63 & 4.50 $\times$ 1.75 \\
Beam P.A. [deg] & 149.3              & 48.5               & 152.3              & 139.7              & 154.0 \\
\hline
\end{tabular}
\end{table*}


\section{Observations}
\label{Sec:observations}

Complementing the observation reported by \citet{Sanchez-Bermudez_2014b}, new single snapshots of Herschel\,36\,A were obtained with AMBER-VLTI on August 10th, 2017, April 28th and August 26th, 2018, as part of our AMBER-VLTI monitoring program \footnote{This project presents observations with the Very Large Telescope Interferometer for the ESO program 597.D-0727(C)}. The observations were conducted in low-resolution mode (R$\sim$35) using only combinations of the VLT unit telescopes (UTs) providing the longest baselines and the highest sensitivity. The data were obtained simultaneously in the $J$, $H$ and $K$ bands, following a sequence of three observations: calibrator (HD\,165920), target, calibrator \citep[which was selected using SearchCal;][]{Bonneau_2006, Bonneau_2011}. Unfortunately, the observations suffered from a variable interferometric fringe tracking performance and the stability of the calibration sequence was negatively affected, especially in the shorter wavelength channels $J$ and $H$, where the phase variance is larger than in the $K-$band, resulting in a larger variance of the fringe tracker offset (called piston). For this reason, the $J$ and $H$ bands could not be properly calibrated and we restricted our analysis only to the $K-$band. 

We reduced the data with \textit{amdlib v3} \citep{AMBER_Tatulli_2007, AMBER_Chelli_2009} to extract the interferometric observables (squared visibilities and closure phases). To keep consistency between the analysis of the new data sets and the one taken on April 17th, 2014, all data sets were reduced using the same constraints. Frames deteriorated by variable atmospheric conditions and technical problems were discarded if any of the following three criteria applied: (a) a baseline flux of less than ten times the noise; (b) a piston larger than 5 $\mu$m and; (c) a visibility signal-to-noise ratio (SNR) amongst the 90\% of the frames with the lowest SNR. Table\,\ref{tab:Obs} reports the key characteristics of the observations. The tracking performance as measured by the root-mean-square (RMS) of the residual phase variation was not always the same for the first and second observation of the calibrator. To avoid systematic calibration errors, we discarded calibrator observations with lower fringe-tracker performance, which led us to select only the second, first, and again the second calibrator observation for the first three epochs, respectively. Furthermore, we noted that tracking was systematically better for the calibrator ($K = 6.0$) than the science target ($K = 6.9$). This caused the overall level of the visibility amplitudes of the science target to be lower, which led us to add a resolved uncorrelated-flux component to our models to avoid overestimating the sizes of the stellar disks (which are expected to be unresolved given the stellar types and their distance). Figure\,\ref{fig:model-fit} in the Appendix shows the visibility data of Herschel 36 obtained with AMBER. Except for two epochs with closure phases which display a 180$^{\circ}$ jump indicative of the detection of two equal-magnitude stellar components, the other phases are consistent with zero. 

\begin{figure*}
\includegraphics[width=\textwidth]{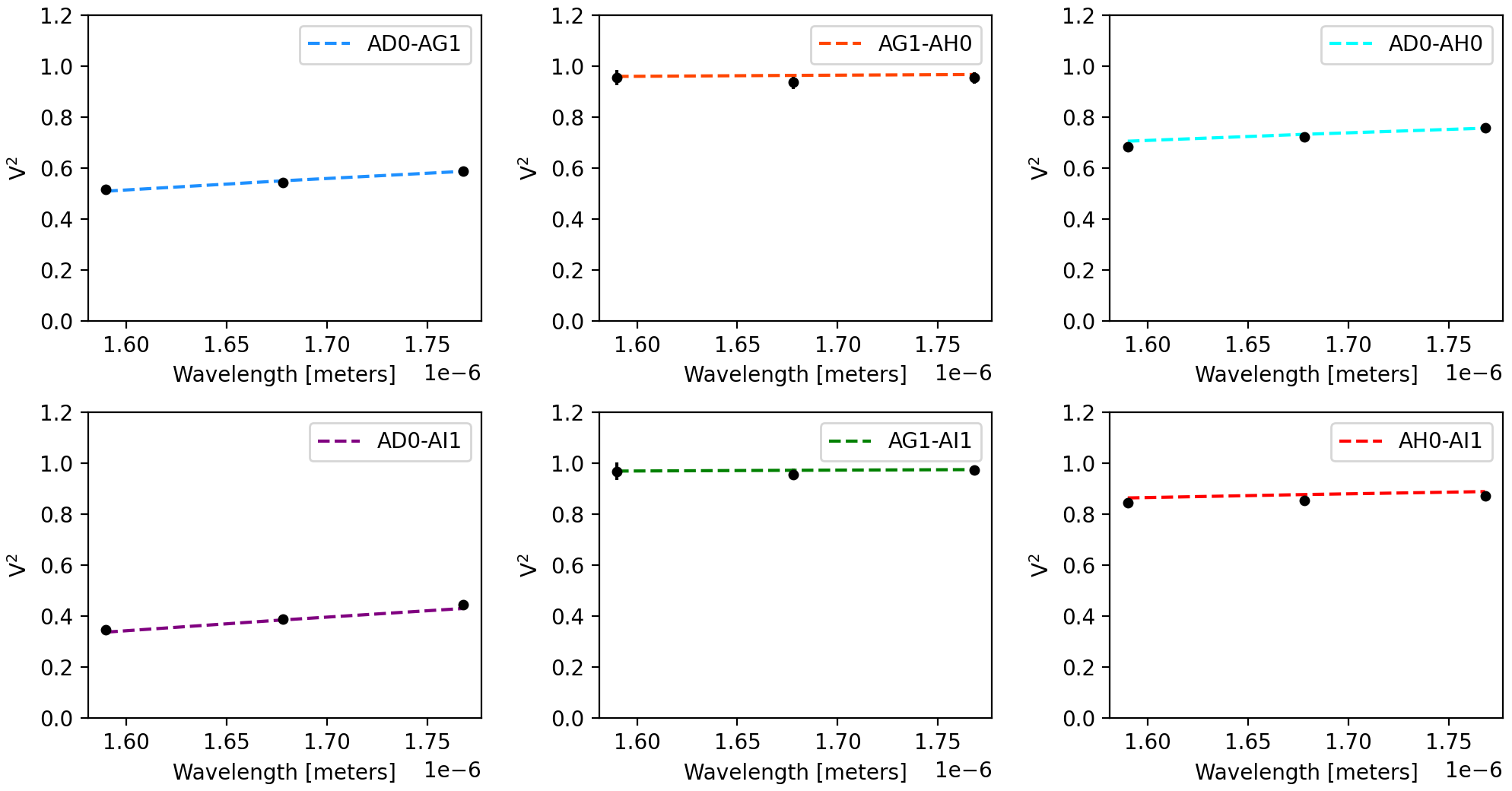}
\caption{Best-fit binary model of the $V^2$ data obtained with PIONIER in H-band.}
\label{fig:model-fit-pionier}
\end{figure*}

Additionally to our AMBER data, one observation from September 3rd, 2014, with the $H-$band PIONIER  instrument of the VLTI was retrieved from the public ESO archive and reduced with the PNDRS pipeline \citep{Le-Bouquin_2011}. PIONIER recorded fringes in three channels across the H-band (Figure\,\ref{fig:model-fit-pionier}). The calibration was also performed with PNDRS, which computes a time-dependent transfer function based on all calibrators observed in the same night, and by which the reduced visibilities of the science targets are divided. The companion was detected as the squared visibilities have values as low as $\sim$ 0.35. All measured closure phases in this data set were consistent with zero. Fig. \ref{fig:model-fit-pionier} in the Appendix displays the V$^2$ data for each one of the six baselines. 


\section{Analysis and Results}
\label{Sec:analysis}

\subsection{Parametric Model Fitting \label{Sec:parametric_model}}

Confirming our 2014 results, we could not resolve the close/spectroscopic binary (Ab1-Ab2) with our new observations. Furthermore, we did not expect to resolve the stellar disks of the components either when taking into account typical O-star diameters seen from a distance of $\sim$1.23 kpc. For example, an O-star of $\sim$ 58 M$_{\odot}$ has a radius of R $\sim$ 14 R$_{\odot}$ \citep{Martins_2005} or R $\sim$ 0.085 mas at the distance of the target, a size quite beyond the resolving power of our interferometer. Therefore, a geometrical model of three point-like objects was used to describe the system Aa-(Ab1-Ab2) at each observational epoch. For the outer orbit, the model considered the system Ab1-Ab2 and the tertiary Aa with normalized fluxes $F_{\mathrm{Ab1Ab2}}$ and $F_{\mathrm{Aa}}$, respectively, separated by a given angular distance and orientation. As mentioned in the previous section, a resolved uncorrelated-flux component $F_{\mathrm{over}}$ was also included to account for the drop in the visibility value due to the variable tracking performance at the time of the observations. The mathematical formulation of our model to compute the complex visibilities, $V(u,v)$, is the following:

\begin{equation}
V(u,v) = \frac{1+F_{\mathrm{Aa}}/F_{\mathrm{Ab1Ab2}}\times e^{-2\pi j (\Delta x u + \Delta y v)}}
              {1+F_{\mathrm{Aa}}/F_{\mathrm{Ab1Ab2}}+F_{\mathrm{over}}/F_{\mathrm{Ab1Ab2}}}\,
\end{equation}

where $u$ and $v$ are the spatial frequencies sampled with our interferometer. Notice that, in this model, the system Ab1+Ab2 is assumed to be at the phase reference. The flux ratio $F_{\mathrm{Aa}}/F_{\mathrm{Ab1Ab2}}$ was assumed to be constant with time and over the band-pass of each observing epoch, which is a reasonable approach. For the minimization, we constrained $F_{\mathrm{Ab1Ab2}}$+$F_{\mathrm{Aa}}$+$F_{\mathrm{over}}$ = 1.0

The model fitting was performed using the OYSTER\footnote{\url{http://www.eso.org/~chummel/oyster}} software. To determine the geometrical parameters of the binary, four parameters were first fitted simultaneously to the visibility data of a given night: $F_{\mathrm{Aa}}/F_{\mathrm{Ab1Ab2}}$, $F_{\mathrm{over}}/F_{\mathrm{Ab1Ab2}}$, $\rho$ and $\theta$. Table\,\ref{tab:binaryfits} displays the best-fit parameters. For completeness, Figs.\,\ref{fig:model-fit} and \ref{fig:model-fit-pionier} in the Appendix show the best-fit model over-plotted on the V$^2$ and closure phases of each one of the different epochs. 
From our modeling, we found consistency with our results reported in \citet{Sanchez-Bermudez_2014b}. The flux ratio between the Aa component and the Ab pair is close to unity for all epochs. It is interesting to mention that only two of the AMBER epochs show a clear jump between 0$^{\circ}$ and 180$^{\circ}$. The rest of the data, including the PIONIER data set, show closure phases at 0$^{\circ}$. This effect is expected because we have a binary with components of (near) equal brightness. Nevertheless, for the epochs with closure phases at zero, it is not possible to break the degeneracy of 180$^{\circ}$ in the position of the Aa component. Only when using a dynamical model as described below we can break this degeneracy (see Sect.\,\ref{sec:orbit}).


\begin{table*}
\caption[]{Herschel\,36\,A best-fit binary parameters}
\label{tab:binaryfits}
\centering
\begin{tabular}{l c c c c c}     
\hline
                            &Apr. 2014&Sep. 2014&Aug. 2017&Apr. 2018&Aug. 2018\\
\hline
$F_{\mathrm{Aa}}/F_{\mathrm{Ab1Ab2}}$ & 0.98      & 1.0     & 1.0  & 1.0  & 0.98 \\
$F_{\mathrm{over}}/F_{\mathrm{Ab1Ab2}}$ & 0.49      & 0     & 0.20  & 0.07  & 0.30 \\
$\rho$ [mas]                       & 1.90      & 1.24  & 2.21  & 1.23  & 2.29 \\
$\theta$ [deg.]                     & 221.3     & 294.8 & 50.3  & 191.0 & 272.6\\
\hline
\textbf{Error ellipse:} & & & & & \\
\hline
Major axis [mas]                        & 0.26      & 0.36  & 0.19  & 0.35  & 0.42 \\
Minor axis [mas]                         & 0.062     & 0.19  & 0.05  & 0.07  & 0.13 \\
Position angle [deg.]                    & 149.3     & 31.4  & 151.8 & 136.2 & 154.3\\
\hline
\end{tabular}
\end{table*}




\subsection{Orbit Aa-Ab}
\label{sec:orbit}

With the new interferometric epochs, initial values for the full set of the outer orbital parameters (and the magnitude difference between components Aa and Ab) were determined by a simultaneous fit of a Keplerian model to the visibility data and the radial velocity (RV) measurements available in the literature. 
Initially, we adopted the orbital values of the Campbell elements $\omega$, $e$, $P$, and $T_0$ from \citet{Campillay_2019}, and then we constrained the Campbell elements $\Omega$, $a$ and $i$ with the interferometric data. Finally, we carried out a full fit of all elements and the masses of the Aa and Ab components to the radial velocities (Aa and Ab1-Ab2) and visibilities, reducing the weight of the latter per measurement due to their number being almost 4 times larger than the number of RV measurements (see also next paragraph). We paid special attention to the fitting of the closure phase (an observable which is quite robust to calibration errors) jumps in the first and last AMBER epochs, since these epochs show clear cosine signatures in the closure phases. Including the RV data into the global fit for the orbit solution allowed us to break the degeneracy in the position of the secondary for those interferometric epochs for which the closure phase values are zero.

The astrometric uncertainty of each of the orbital positions fitted to the interferometric data is an ellipse derived from the synthesized beam and thus depends on the actual $u-v$ coverage delivered by the aperture-synthesis observation. Due to the possible presence of systematic errors related to the calibration, which may also lead to correlations between visibilities of neighboring spectral channels, we conservatively adopted a full-spectral correlation factor of 15 (number of channels for AMBER), thus lowering the number of independent data points by the same factor. Furthermore, we chose a conservative $5\sigma$ confidence level for the contour in the $\chi^2$ surface to which we fit the uncertainty ellipse. 

\begin{figure}
  \centering
  \includegraphics[width=\linewidth]{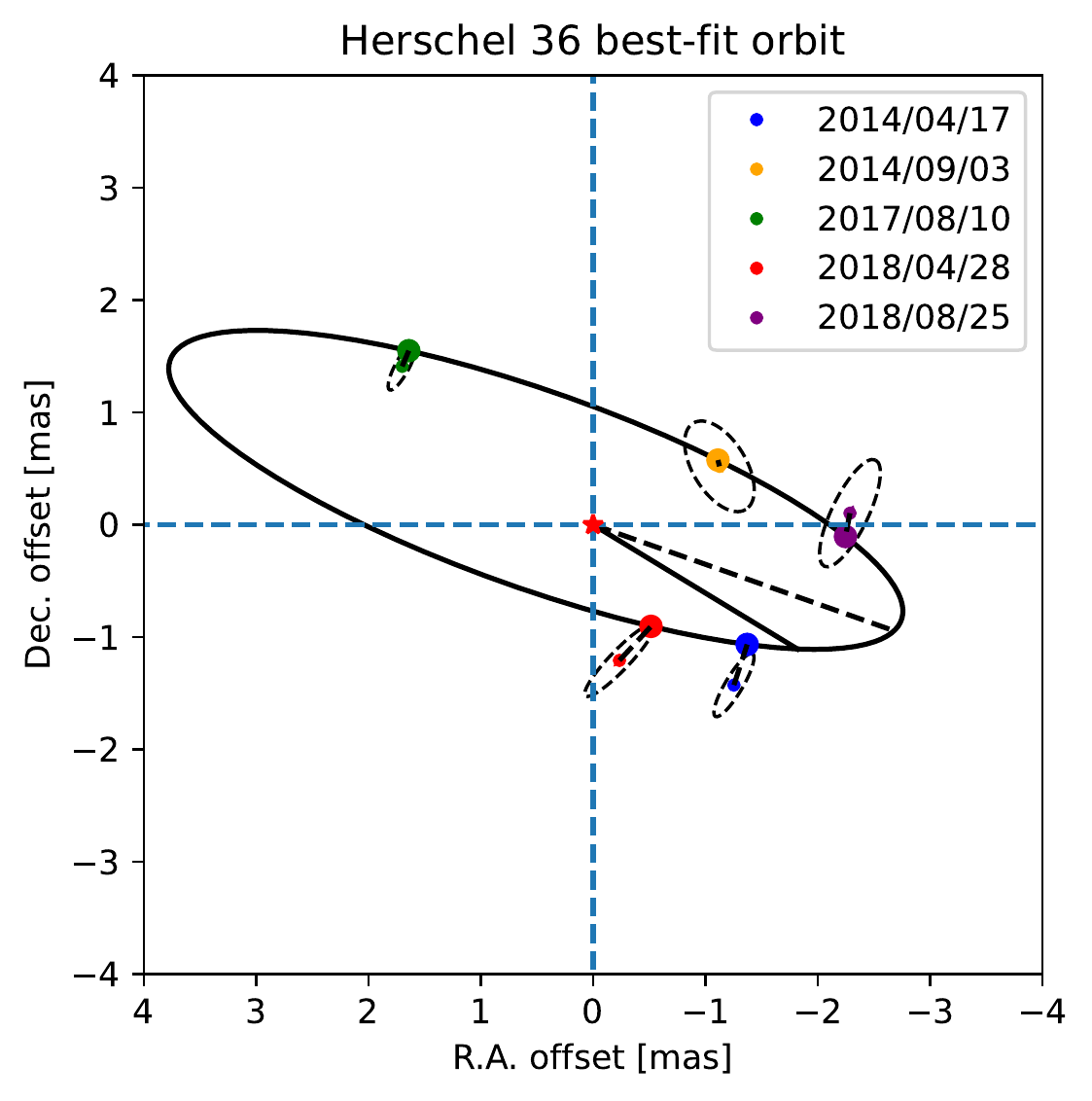}

\caption{Best-fit common orbital solution of Aa around the system Ab1-Ab2 (indicated with a red star at the center of the plot), over-plotted on the astrometric error ellipses with dashed lines connecting their center to the position predicted by the orbit model. The solid straight line indicates the periastron and the dashed straight line indicates the position angle of the ascending node. The different interferometric epochs are plotted with different colors (see labels on the figure).}

\label{fig:orbit_psn}
\end{figure}

\begin{figure}
  \centering
  \includegraphics[width=\linewidth]{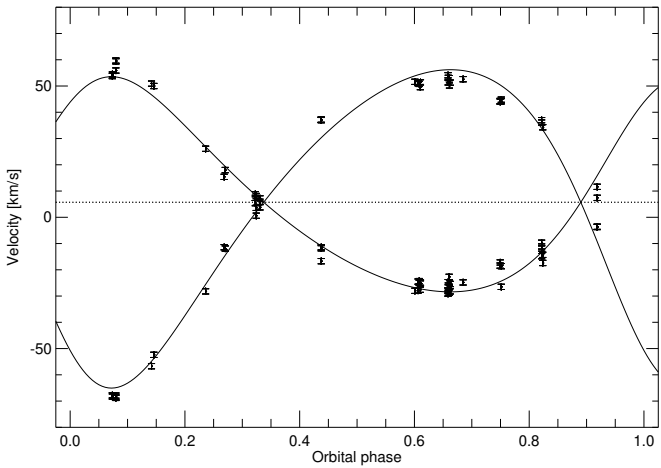}
  \caption{Best-fit common dynamical solution of Aa around the system Ab1-Ab2, over-plotted on the RV data from \citet{Campillay_2019} for the Aa-Ab orbit.}  
\label{fig:orbit_vel}
\end{figure}

 
Table \ref{tab:Best-fit-orbit} shows the results with the best-fit parameters of the common orbit, while Figs.\,\ref{fig:orbit_psn} and \ref{fig:orbit_vel} display the best-fit orbit plotted over the astrometric positions of the Aa-Ab component and the radial velocities, respectively. 



\begin{table}
\caption[]{Herschel\,36\,Aa-Ab best-fit orbital parameters and masses}
\label{tab:Best-fit-orbit}
\centering
\begin{tabular}{l c } 
\hline
Parameter & Value  \\
\hline
$\omega^1$ [deg] & 139.8 $\pm$ 3.8 \\
$e$ & 0.219 $\pm$ 0.014 \\
$P$ [d] & 496.82 $\pm$ 0.78 \\
$T_0$ [HJD] & 2456278.8 $\pm$ 6.5 \\
$a$ [mas] &  3.49 $\pm$ 1.13\\
$\Omega$ [deg] & 250.8 $\pm$ 5.7 \\
$i$ [deg]$^2$ & 75.3 $\pm$ 4.8 \\
Mass Aa & $22.3 M_{\odot} \pm 1.7$ \\
Mass Ab & $33.0 M_{\odot} \pm 2.4$ \\
$\gamma$ Vel. [km/s] & 5.72 $\pm$ 0.46 \\
\hline
\end{tabular}
\begin{list} {}{} \itemsep1pt \parskip0pt \parsep0pt \footnotesize
      \small
      \item $^1$ The angle of the periastron passage reported corresponds to the one of the system Ab orbiting the component Aa. To reproduce the orbit in Fig. \ref{fig:orbit_psn}, 180$^{\circ}$ should be added to the reported value
      \item $^2$ Angle measured from the plane of the sky to the plane of the orbit
    \end{list}
\end{table}

\subsection{Orbit Ab1-Ab2}
\label{sec:orbit_ab1-ab2}

Even though the inner orbit cannot be resolved with our observations, its semi-major axis, inclination and, thus, the stellar masses of components Ab1 and Ab2 can be computed from the Campbell elements $\omega$, $e$, $P$ and $T_0$ already known from the analysis of \citet{Campillay_2019} and by imposing the condition that the orbital parallaxes\footnote{The orbital parallax corresponds to a distance derived from the total mass of the system, the period and angular semi-major axis of the orbit} of the orbits Aa-Ab and Ab1-Ab2 must be identical for a dynamically interacting hierarchical triple system. 

To obtain unconstrained orbital elements of the inner orbit Ab1-Ab2, first, we used the mass ratio $M_{\rm Ab2/Ab1}=0.613$, derived from \citep{Campillay_2019}, and the total mass of Ab, obtained from our interferometric data and reported on Table \ref{tab:Best-fit-orbit}. With these data, we get $M_{\rm Ab1}=20.46\pm1.49$ and $M_{\rm Ab2}=12.54\pm0.91$. The orbital parallax of the outer orbit is $\pi_{\rm Aa-Ab}=0.75\pm0.24$ mas. Thus, we used this value as the parallax of the inner orbit, since the spectroscopic binary and the tertiary component are gravitationally bound. Therefore, we could get the best-fit value of the semi-major axis of the inner binary, resulting in $a_{\rm Ab1-Ab2}=0.0629 \pm 0.02$ mas.

Finally, we also fitted the inclination of the inner orbit to reproduce the radial velocity measurements for components Ab1 and Ab2, our best-fit value results in $i=53.7^\circ\pm2.1^\circ$ (the uncertainty reported is related to the error bars of the semi-amplitudes). Since this compact system is expected to have zero eccentricity, the angle of the periastron passage $\omega$=0. Also, since we are not able to resolve the components, the angle of the ascending node, $\Omega$, is the only element which remains unconstrained.

\begin{figure}
  \centering
  \includegraphics[width=\linewidth]{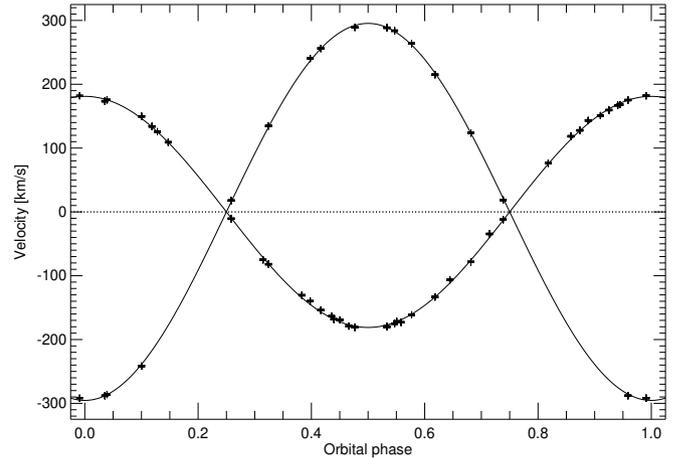}
  \caption{Best-fit dynamical solution for the radial velocities of Ab2 and Ab1, over-plotted on the RV data from \citet{Campillay_2019} for the Ab1-Ab2 orbit.}  
\label{fig:orbit_vel_ab}
\end{figure}

\section{Discussion}
\label{Sec:discussion}

In this paper, we have analyzed data from three new observations with AMBER and from one archival PIONIER observation of the massive triple system Herschel 36\,A. Combined with our previous astrometric solution reported in \citet{Sanchez-Bermudez_2014b}, we were able, for the first time, to determine the elements of the outer orbit Ab1-Ab2, as well as those of the inner orbit. Herschel 36\,A has an interesting configuration which is not common in other triple hierarchical systems \citep[for example HD\,150136;][]{Sanchez-Bermudez_2013,Mahy_2018} in which the most massive and brightest component is a member of the close binary; for this target it is not the case. 

\textit{On the mass of the target:} 
The derived mass of component Aa is consistent within 1-$\sigma$ \citep[following the observational calibration of ][]{Martins_2005} with the spectral classification of an O7.5 V star. However, notice that the calibrations of \citet{Martins_2005} were derived for stars on the main-sequence while the components of Herschel\,36\,A might still be on the zero-age main-sequence. The mass derived for component Ab1 is somewhat higher than the value quoted in the calibration by \citet{Martins_2005} for an O9.5 V star, but still in agreement given the large uncertainties in the mass estimate (1$\sigma \sim$2 M$_{\odot}$). Finally, Ab2 is in agreement with the observed mass for a star with similar spectral type \citep[see e.g., the  B0.5 V star HD\,315031;][]{Gonzalez_2014}. With the new estimate of $a_\mathrm{Aa-Ab}$=3.49$\pm$1.13 mas and the total mass derived from the orbital solution (see Table \ref{tab:Best-fit-orbit}), we estimated a distance to the target of d = 1.33 $\pm$ 0.32 kpc. This value is consistent within 1-$\sigma$ with the estimates reported by \citet{Campillay_2019}, \citet{Maiz-Apellaniz_2022} and with the GAIA EDR3 parallax of Herschel 36 being $0.775 +/- 0.024$ mas. More precise constraints on the distance to the system could be obtained by refining the orbital estimation with better astrometric VLTI observations obtained through beam-combiners like GRAVITY.


\textit{On the coplanarity of the orbits:} It is suspected that the formation of massive stars happens in dense environments at the core of molecular clouds. As mentioned before, two main classes of theories are contemporary to describe massive star formation: \textit{Competitive Accretion} \citep{Bonnell_2001} and \textit{Core Accretion} \citep{Tan_2014}. In the latter one, the formation of high-mass star(s) is subject to the existence of a self-gravitating clump which collapses into an accretion disk where the forming star(s) gain their mass. In \textit{Competitive Accretion}, the material is drawn from a chaotic environment at scales larger than the typical size of massive clumps ($\sim$0.3 pc) and the formation of the massive star(s) is not subject to the presence of a massive contracting pre-stellar clump.

In \textit{Core Accretion}, the formation of binaries, or multiple systems, happens due to gravitational instabilities in the accretion disk \citep{Krumholz_2007d}. This scenario favors coplanar orbits between different stellar companions. \citet{Kratter_2007} show that a binary system as massive as $\sim$10$^2$ M$_{\odot}$ could be formed within a massive disk with a relative mass fraction $\mu$ = M$_{\mathrm{disk}}$/(M$_{\mathrm{disk}}$ + M$_*$) = 0.5 which undergoes a phase of local instability. However, the global disk structure remains stable for accretion onto the stellar companions to continue. On the other hand, the turbulent environment proposed for the  \textit{Competitive Accretion} theory sets the initial conditions for early dynamical interactions \citep[see e.g.,][]{Larwood_1997, Bonnell_2003} which favor the formation of non-coplanar systems.

We were able to derive the inclination of the inner orbit (i$_{\mathrm{Ab1-Ab2}}$ = 75.3$^{\circ} \pm$4.8$^{\circ}$) and found it to be tilted relative to the outer orbit (i$_{\mathrm{Aa-Ab}}$ = 53.7$^{\circ} \pm$2.1$^{\circ}$) by $21.6^{\circ} \pm$ 5.2$^{\circ}$. This small tilt between the two orbits suggests that the target could have been formed via \textit{Core Accretion}. Nevertheless, as the value of the mean inclination difference is not quite at 5-$\sigma$ based on the reported error, {\it we cannot completely rule-out \textit{Competitive Accretion} as the formation mechanism for our target}. To settle better constraints on its formation, dedicated hydrodynamic and N-body simulations must be done. For example,  for the \textit{Competitive Accretion} scenario it would be interesting to quantify (i) the particular set of initial conditions for the dynamical interaction to keep the system long-term stable, and (ii) the process of migration of the different components to harden the inner binary system (probably created from the same core) and drag-out the outer component. For the \textit{Core Accretion}  it is necessary to quantify the timescale of the fragmentation and accretion processes. These analyses should certainly be considered for future reports on the target. Furthermore, future observations with PIONIER-VLTI and/or GRAVITY-VLTI are envisioned to refine the orbital estimates and to have better initial parameters for more accurate simulations.


\section{Summary and Outlook }
\label{sec:summary}
In this paper, we present new interferometric observations of the triple system Herschel\,36\,A. From the new information derived, in combination with previously published spectroscopic data, we inferred (i) the total mass of the system, (ii) the orbital parameters of the outer orbit and (iii) the difference between the planes of the inner and outer orbit of the triple system. A refinement of the orbital solution is necessary to better constrain the physical parameters derived. Therefore, new interferometric observations should be requested in the near future  for this purpose. From the coplanarity analysis, it is not conclusive whether the system formed from the collapse of a single core or in a competitive accretion environment. Detailed numerical simulations may be useful to discriminate one scenario from the other. Finally, to better understand the formation of high-mass multiple systems, systematic analyses like this must be extended to other targets, for example the triple $\theta^1$ Ori B$_{1,5,6}$ \citep[see][]{Karl_2018} which has a geometry similar to Herschel 36\,A.

\section*{Acknowledgements}
All authors thank the anonymous referee for his/her constructive comments to improve the present work. R.S. and A.A. acknowledge financial support from the State Agency for Research of the Spanish MCIU through the "Center of Excellence Severo Ochoa" award for the Instituto de Astrofísica de Andalucía (SEV-2017-0709). R.S. acknowledges financial support from national project PGC2018-095049-B-C21 (MCIU/AEI/FEDER, UE).  A.A. acknowledges financial support from national project PID2020-117404GB-C21 (MCIU/AEI/FEDER, UE). J.S.B. acknowledges the financial support from the "Visitor Scientist Program" of the "Center of Excellence Severo Ochoa" provided by the IAA-CSIC; and to the "ESO-Garching Visitor Program" of the European Southern Observatory. This work presents results from the European Space Agency (ESA) space mission Gaia. Gaia data are being processed by the Gaia Data Processing and Analysis Consortium (DPAC). Funding for the DPAC is provided by national institutions, in particular the institutions participating in the Gaia MultiLateral Agreement (MLA). The Gaia mission website is https://www.cosmos.esa.int/gaia. The Gaia archive website is https://archives.esac.esa.int/gaia. 

\section*{Data availability}
The new interferometric data underlying this article will be shared on reasonable request to the corresponding author.



\bibliographystyle{mnras}




\appendix

\section{Interferometric observables and their best-fit model}

\begin{figure*}
\includegraphics[width=1.0\textwidth]{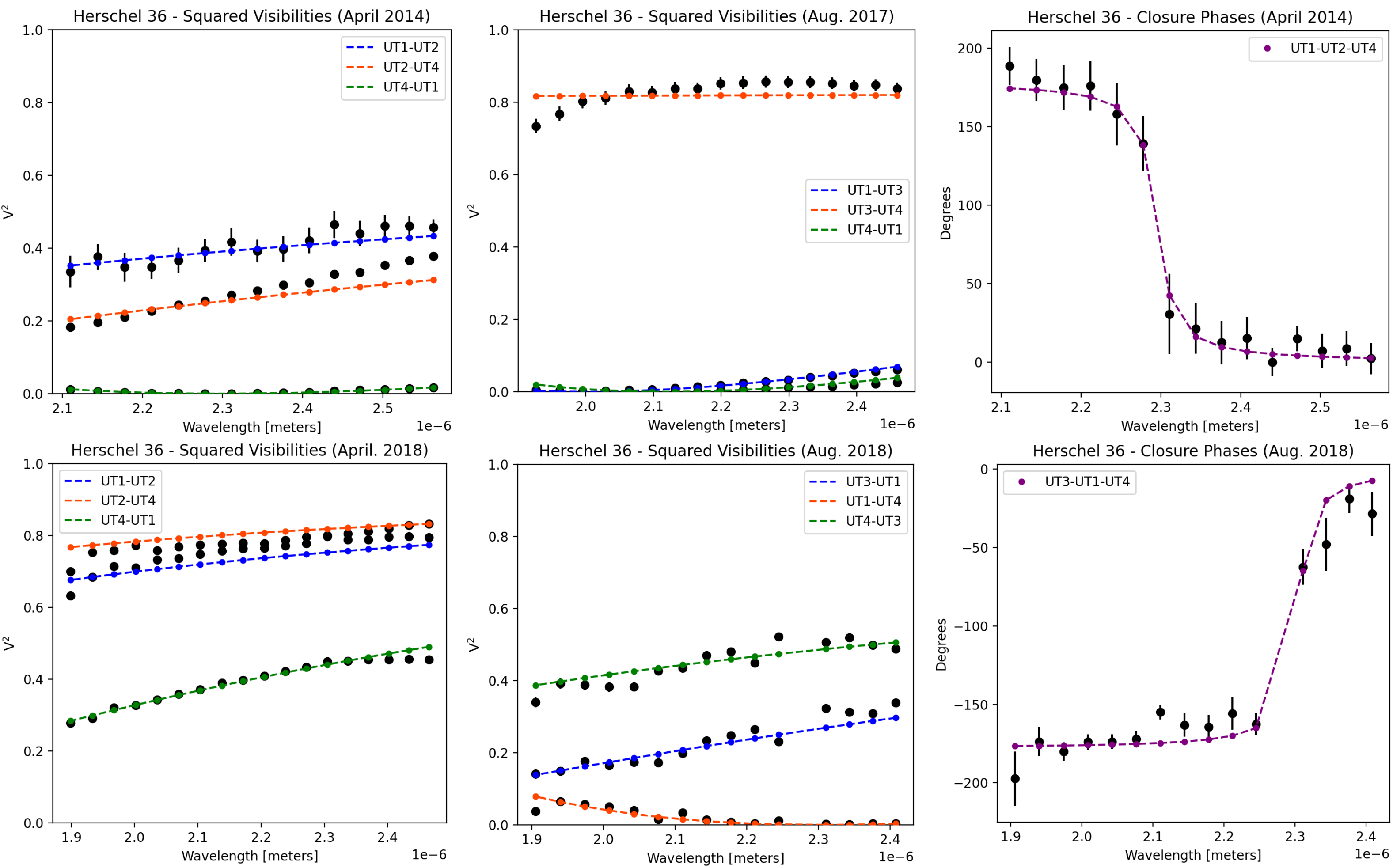}
\caption{The first and second column show the best-fit binary models to the $V^2$ data obtained from the four epochs of AMBER-VLTI data reported; the last column show the best-fit binary model to the closure phases in the data which shows a (cosine) signature different from zero. Panels display with black dots the interferometric data, while the best-fit model is plotted in color for different baselines (see legends in the plots).}
\label{fig:model-fit}
\end{figure*}

\bsp	
\label{lastpage}
\end{document}